# Statistical Methods and Workflow for Analyzing Human Metabolomics Data

Joseph Antonelli,*[1,2] Brian L. Claggett,*[1] Mir Henglin,[1] Jeramie D. Watrous,[3] Kim A. Lehmann,[3] Pavel V. Hushcha,[1] Olga V. Demler,[4] Samia Mora,[1,4] Teemu J. Niiranen,[5] Alexandre C. Pereira,[6] Mohit Jain,*[3] Susan Cheng*[1,5]

*Equal author contribution

[1]Cardiovascular Division, Brigham and Women's Hospital, Harvard Medical School, Boston, MA

[2]Department of Biostatistics, Harvard T.H. Chan School of Public Health, Boston, MA

[3]Departments of Medicine & Pharmacology, University of California San Diego, La Jolla, CA

[4]Preventive Medicine, Brigham and Women's Hospital, Harvard Medical School, Boston, MA

[5]Framingham Heart Study, Framingham, MA

[6]Department of Genetics, Harvard Medical School, Boston, MA

**Correspondence:** Mohit Jain, MD, PhD, University of California, San Diego, mjain@ucsd.edu; and, Susan Cheng, MD, MPH, Brigham and Women's Hospital, Harvard Medical School, scheng@rics.bwh.harvard.edu.

# ABSTRACT

High-throughput metabolomics investigations, when conducted in large human cohorts, represent a potentially powerful tool for elucidating the biochemical diversity underlying human health and disease. Large-scale metabolomics data sources, generated using either targeted and untargeted platforms, are becoming more common. Appropriate statistical analysis of these complex high-dimensional data will be critical for extracting meaningful results from such large-scale human metabolomics studies. Therefore, we consider the statistical analytical approaches that have been employed in prior human metabolomics studies. Based on the lessons learned and collective experience to date in the field, we offer a step-by-step framework for pursuing statistical analyses of human metabolomics data. We discuss the range of options and potential approaches that may be employed at each stage of data management, analysis, and interpretation and offer guidance on the analytical decisions that need to be considered over the course of implementing an analysis workflow. Certain pervasive analytical challenges facing the field warrant ongoing focused research. Addressing these challenges, particularly those related to analyzing human metabolomics data, will allow for more standardization of as well as advances in how research in the field is practiced. In turn, such major analytical advances will lead to substantial improvements in the overall contributions of human metabolomics investigations.

**Introduction**

Rapid advancements in mass spectrometry (MS) technologies have enabled the generation of large-scale metabolomics data in human studies. These technical advances have outpaced the development of statistical methods for handling and analyzing datasets of now burgeoning size and complexity.[1-3] Early investigations using metabolomics technologies have applied a variety of statistical methods in analyses of datasets containing up to 200 metabolite measures, typically acquired from a targeted metabolomics platform collected from human studies involving tens to hundreds of observations. Available untargeted metabolomics platforms now allow the measurement of thousands of metabolite variables, most which are unknown molecular species that demonstrate varying levels of intercorrelations within a given dataset as well as correlations with a given clinical outcome. Furthermore, current metabolomics technologies have augmented throughput capacity that can facilitate data collection for thousands of observations per human cohort experiment. At present, however, there are no existing standard protocols for analyzing these increasingly complex metabolomics data. Thus, herein, we review the available statistical methods for analyzing high-dimensional metabolomics data in the setting of a clinical study. We also outline an accessible yet flexible approach that can be used to optimize sensitivity and specificity for identifying potentially important metabolites associated with a clinically relevant outcome.

**Analytical Challenges**

Due to the increasing complexity of metabolomics data, combined with the variety of different study designs employed in practice, customized approaches are often required for analyzing metabolite variation in relation to clinical outcomes. Notwithstanding differences in study design and data structure, common to almost all metabolomics datasets is the need to address certain statistical considerations (**Table 1**). An initial key consideration is missingness, given that all metabolomic data invariably demonstrate patterns of missing values that are often but not always more frequent for low abundant metabolites. Contributing factors include underlying biology (e.g., very low or true zero values due to biological differences between healthy and disease-enriched cohorts) and technical issues (e.g., actual values below the detection limits for a given method, which may or may not be rectified using different or complementary methods). The etiology of missing values will vary, at least in part, based on the platform used to profile metabolites (i.e. NMR, GC-MS, LC-MS). In rare circumstances, for certain types of metabolites (e.g., derivatives of known toxic exposures), missing values may be most appropriately coded as true zero values as opposed to imputed. For non-zero metabolite variables, transformation is usually recommended due to frequently right-skewed distributions. Irrespective of distribution patterns, metabolite variables always demonstrate intercorrelations. Importantly, the extent to which intercorrelations exist between metabolites will vary substantially between datasets due to a variety of factors, including those related to study sampling as well as technical issues. Furthermore, even within a given study sample, metabolite variation will be influenced by time-dependent factors given that a portion of the human metabolome changes dynamically in response to acute perturbation or stress while another portion of the metabolome exhibits relatively little change over time, except in response to major chronic exposures.

**Statistical Methods for Analyzing Metabolomics Data**

A variety of different statistical methods are available for analyzing high-dimensional metabolomics data. Methods that have been applied in prior or ongoing clinical metabolomics studies are summarized in **Table 2** and are also described in more detail herein.

Univariate analyses with multiple testing correction, such as the Bonferroni correction for controlling the global type I error rate or the Benjamini-Hochberg correction for controlling the false discovery rate (FDR),[1] have previously been applied in a variety of predominantly targeted metabolomics studies.[2-4] This approach involves M tests, where M is the total number of metabolite variables analyzed separately in relation to an outcome of interest. The P value for each separate metabolite test can be considered significant or non-significant based on a P value threshold that is corrected to account for the fact that multiple hypotheses have are being tested. The Bonferroni method is commonly used and deems a metabolite significant if it is smaller than 0.05/M, while other corrections can be used to control the false discovery rate. Overall, univariate analysis with multiple testing correction is an attractive approach because it is simple to implement and provides a measure of statistical significance for each covariate that is easy to interpret. However, this approach alone does not account for associations between metabolites and conditional associations. For instance, a given metabolite may appear significantly associated with an outcome in isolation but does not demonstrate a significant association when other metabolite associations are taken into account. Furthermore, approaches to account for multiple testing, such as the Bonferroni correction or even FDR, are considered conservative in the setting of a large number of analyzed metabolites, leading to limited statistical power overall.

The principle components analysis (PCA) approach[5] is designed to reduce the dimension of the number of metabolites being analyzed, assuming there is substantial correlation between metabolites in a given dataset. Thus, PCA directly addresses the issue of intercorrelatedness and has been used in prior human metabolomics studies, usually in combination with other methods. The PCA approach takes the original metabolite factors and finds linear combinations of these factors that are orthogonal to each other and that explain the most variation in the metabolite dataset. The PCA approach can be used to identify factor combinations associated with a given outcome, but PCA is not intrinsically designed to identify original predictor variables (i.e., metabolites) of importance. A varimax rotation can be used to increase the interpretability of principal components, given that principal components are comprised of a small subset of the original metabolites.[6] Those metabolites contributing to a given principal component can be ranked by the importance of their contributions. However, for each contributing metabolite, a measure of the magnitude of its association with given outcome is not provided, and a test of the significance of association is also not provided.

The partial least squares (PLS) regression method[7] aims to maximize the covariance between a matrix of metabolites and a continuous outcome (or categorical outcome using the PLS discriminant analysis [DA] variation[8]) by decomposing metabolite and outcomes data into latent structures. This approach aims to maximize the covariance between the outcome and matrix of metabolites by projecting both to linear subspaces of the original variables. While finding reduced dimensions that can explain the outcome variable, PLS regression generally only provides a measure of variable importance and does not naturally perform variable selection, though a number of ad hoc approaches for variable selection have been proposed.[9,10] Nonetheless, no clear-cut approach is best for determining which metabolites are actually important in predicting the outcome. Sparse extensions of both approaches have been proposed,[11-15] which add a penalty to the loading scores forcing some of the variables to have zero weight in the final model. One can take the list of non-zero metabolites to be the metabolites that are deemed important. This is potentially a very fruitful direction for identifying important metabolites.

The linear discriminant analysis (LDA) approach aims to find linear combinations of metabolite variables that are best able to separate classes of a categorical outcome. Although LDA cannot be used for a continuous outcome, such an outcome may be discretized for LDA application. Conventionally, LDA will perform poorly or can even fail completely when the number of covariates exceeds the number of subjects,[16,17] a common feature of metabolomics datasets. Furthermore, the LDA approach does not intrinsically identify a set of important variables and can only be used to create variable rankings or importance measures. For this reason, we consider a sparse version of LDA,[18-20] which again adds a penalty for the variable loadings and, thus, allows for simultaneous variable selection. As with PCA and PLS, LDA can provide a measure of variable importance that can be used for metabolite ranking.

The least absolute shrinkage and selection operator (LASSO) approach aims to fit a model that regresses the outcome against all of the metabolites simultaneously and applies a penalty to the magnitude of regression coefficients to achieve sparse variable selection.[21] Generally, regression models are extremely noisy, if not infeasible, when the number of metabolites is large. LASSO applies a penalty to the magnitude of the regression coefficients, which forces the regression coefficients for many coefficients to be zero while also shrinking others in magnitude. LASSO is useful in both prediction and metabolite selection. One can take the collection of metabolites with non-zero regression coefficients from LASSO to be "significant" in the sense that they are associated with the outcome. This should not be confused with statistical significance in the general sense of p-values and rejecting null hypotheses; however, it is a powerful tool for variable selection. To perform LASSO, one must also select a tuning parameter for the penalty; however, this is easily done via cross validation and is implemented in statistical software. LASSO has many desirable large sample properties including model selection consistency, which states that one can select the proper metabolites with probability tending toward 1 if we have enough data. LASSO is known, however, to struggle in small samples with highly correlated covariates as it will simply choose one among the group of correlated variables and force the others to be zero. In these cases, related methods such as elastic net, which is a compromise between the LASSO and ridge regression procedures, can be utilized as well. LASSO and its variants can be implemented in the glmnet package in R.[22]

Random forest is a non-parametric ensemble method that prioritizes prediction by attempting to find non-linear patterns in metabolites that can explain variation in a given outcome.[23] Random forests are very powerful tools if the relationships between the metabolites and the outcome are complex and non-linear. One drawback of this approach is that, similar to PCA, it does not provide a measure of statistical significance or provide any p-value or equivalent quantity. Metabolite importance, however, can be assessed by removing metabolites one at a time, re-running the procedure, and seeing how much predictive capability was lost. This provides the analyst with a list ranking the most important metabolites but does not provide a cutoff for which metabolites are significant. Random forests can be implemented in either the randomForest or h2O packages in R.[24,25]

Additionally, there is a vast array of methods in the machine learning literature that provide very flexible models for handling data with a large number of covariates. These approaches are appreciably powerful tools for predicting a given outcome of interest although, in many cases, at the expense of interpretability of resulting models. Examples of commonly used approaches include support vector machines (SVM), neural networks, and the previously mentioned random forests.[26,27] All of these methods are subject to many of the same limitations, as discussed for random forests, in that they can provide a variable importance measure but do not provide a set of variables that can be thought of as statistically significant.

A note regarding sparsity in statistical methods is warranted. A general problem of many approaches that may be used to analyze metabolomics data is that they do not easily result in a final list of "top hit" metabolites associated with a given outcome of interest. Instead, they are useful for providing heuristic measures that indicate variable importance and, in doing so, do not eliminate clinical or other covariates from the model of total covariates included in analyses. For this reason, we recommend focusing on sparse alternatives to statistical methods because, as discussed above, they are intended to directly address this issue. In this context, sparsity is based on the assumption that the number of true positives is limited, such that the contribution to variation in an outcome can be defined by a set number of non-zero values for a set number of coefficients, with the remaining being zero values. One of the naivest ways to achieve a sparse result, for instance, would be a stepwise (e.g., forward) selection, but alternate methods are preferred when the number of metabolites is far in excess to the number of observations.

**Metabolomics Analysis Workflow**

In addition to considering which statistical methods to apply in relating curated metabolomics data to clinical outcomes, a series of study design and data management steps are required as part of a complete analytical workflow. While not intended to be comprehensive, an overview of these steps is provided below.

*Step 1: Study Design*

Prior to beginning any experiment, all aspects of study design should be carefully determined. This process includes a decision on whether metabolite predictors of a given outcome will be investigated in a case-control (and, if so, the number of cases and controls), a case-cohort, or total cohort design. In this step, it is critically important to consider the number of outcome occurrences (i.e., cases) available for analysis and the anticipated effect size for metabolomic variation in relation to case status. Also, it is necessary to decide on whether the metabolomics profiling method to be used will be untargeted or targeted and, if targeted, the type of targeted approach. This decision will determine the approximate number of metabolites that will be measured. Taken together, information on the total number of observations, number of cases, and number of metabolites measured will determine the extent to which the experiment will be adequately powered for detecting clinically significant associations of interest. Similar considerations are needed for analysis of continuous as opposed to binary outcomes. Overall, this step is essential for conducting analysis of metabolomics data wherein the number of metabolites measurable typically far exceeds the number of individuals studied within a given experiment.

*Step 2: Data Management*

It is becoming increasingly recognized that several key data management steps are extremely important for ensuring not only the integrity but also feasibility of metabolomics data analysis. For instance, it is always important to perform a careful assessment for batch-to-batch variability that can often persist even after preprocessing steps have been successfully completed (e.g., for alignment of mass spectral features).[28] Accordingly, an informed decision is possible regarding whether normalization of data to internal standards or to pooled plasma measures is needed.[29-31]

All datasets should be examined for their data structure, including the distribution and type of missingness across metabolites and across individuals. Different approaches to handling missingness may be suitable depending on the types of metabolites profiled (e.g., known to be rare, low abundant, or technically difficult to detect). Investigators have imputed values that are a fraction (e.g. 0.5) the lowest value measured for a given analyte, based on the assumption that most missingness for a given platform is due to limits of detection. Other imputation methods are possible, including replacement of all missing values with zero. Analytes with a large proportion of measures missing may also be treated as dichotomous variables (non-missing versus missing), subset analyses are always possible, and some investigators have elected to exclude analytes with substantial missingness from all analyses. All approaches to handling missing values may introduce bias, depending on the method and cohort characteristics; a deeper examination of the relative merits of each approach is a subject of ongoing research (e.g. sensitivity analyses performed with and without variables potentially requiring imputation). After having issues related to missingness addressed, metabolite variables typically benefit from applying transformation and scaling to allow for appropriateness and comparability of statistical analyses (**Figure 1**). Although natural log transformation is commonly used for most or all metabolite variables in a given dataset, given the typical high proportion of metabolites with right-skewed distribution, this approach may not be optimal for all variables.[32] We recommend performing a natural log transformation of all metabolite variables and assessing the skewness before and after transformation. For those metabolites for which the skewness is not improved after transformation, we recommend retaining the original untransformed variables for further analysis, with transformed values used for the other metabolites. While standardization of transformed variables is also commonly used, so that magnitudes of effect are comparable across models, other approaches such as Pareto or level scaling may be more appropriate for certain study designs (i.e. taking into consideration the main research questions and outcomes of interest).[32]

*Step 3: Optional Simulation Analyses*

If feasible, an ideal approach is to develop a dataset that simulates design and size of the study at hand, including the number of metabolites measured and the expected or known number of outcomes. Simulated data allow for comparisons of performance of the different statistical analysis methods, so that the most optimal method can be selected for a given study. Simulations can be particularly useful for guiding the design of studies that are in the early planning phases, particularly those for which the number of metabolites and the number of study subjects has yet to be determined. To perform simulations that replicate a given study, some features of the data must be known. For instance, if prior data is already available, the empirical distribution of the metabolites can be used for simulation (**Figure 2**). One can resample the rows of the true data to obtain simulated data that closely replicate the true data, and then an outcome can be generated assuming a pre-specified relationship with the metabolites. Another possibility is to learn the covariance matrix among all the metabolites in the study and then draw values from a multivariate normal distribution with this covariance to create simulated metabolite variables. If an example of true data is not available and simulations are being used to design a new study, then a dataset from previously conducted related study may be used to guide the simulation design. With simulated datasets in hand, it is possible to obtain power calculations that are relevant and applicable for planning study design. One can simulate metabolites and an outcome and repeat this process many times over while recording the percentage of times a metabolite marker of interest is identified, which provides the power at the given sample size. This process can be performed iteratively with a number of different sample sizes, and different pre-specified associations between the metabolites and outcome, to gain a better understanding of the power available to identify signals in the planned study.

*Step 4: Cross-Sectional or Prospective Analyses (i.e., Outcomes Analyses)*

After selecting the most appropriate statistical approach, based on prior experience or the results of simulation analyses (Step 3 above), applying a method that includes internal cross-validation needs to be considered. Cross-validation procedures are intended to optimize generalizability and reduce variability of results by performing repeated analyses on different partitions of the dataset and then averaging the results to estimate a final model. For instance, in k-fold cross-validation, the total dataset is randomly divided into k equally sized subsets, and k-1 subsets are analyzed while reserving a different single subset for validation during each iteration. An alternative to cross-validation is conventional validation or dividing the original dataset into training and validation subsets (e.g., 2/3, 1/3) while assuming the original dataset size is large enough to accommodate this approach and a separate validation cohort is not available. Yet another option for assessing generalizability is to apply other statistical methods and compare results,[33] as an investigator can feel more confident about results that are consistently obtained irrespective of the statistical technique used (**Figure 3**).

*Step 5: Visualization of Main Findings*

There are many different approaches to visualizing metabolomics data as part of the workflow for understanding the structure of a dataset as well as interpreting and communicating results of relational analyses. Many approaches are borrowed from other established fields. Common visualization types include Manhattan plots, volcano plots, and heatmaps, each of which offer complementary information with respect to conveying information about the significance or effect size of associations between multiple metabolites and one or more outcomes of interest (**Figure 4**). Because the number of metabolites analyzed in relation to a given clinical trait or outcome may be quite large, visualization methods that include clustering on inter-metabolite associations may be additionally informative.[34] Visualizations that can provide information about network relationships, particularly when combined with knowledge about putative metabolic pathways,[34] may also be relevant depending on the types of metabolites measured. For untargeted metabolomics studies, network analyses based on known or previously reported putative biological relationships will be limited based on the extent to which most newly discovered analytes of interest will novel (i.e. of previously unknown identity).

### *Step 6: Prioritization of Results for Follow-Up Investigations*

When using an untargeted MS method, the majority of significantly associated metabolites will invariably be novel (i.e., previously unidentified molecular species). Thus, an imperative next step in the scientific process is to identify such novel molecules of potential clinical importance. However, the process of identifying the specific chemical structure of a previously unknown small molecule is time-consuming and potentially very resource-intensive, depending on the molecule's relative abundance in available biospecimens and other characteristic features. Therefore, the results of statistical analyses should ideally be robust and convincing before they are used to direct efforts made towards novel small molecule identification. To this end, several statistical approaches to prioritizing small molecules for follow-up identification are possible. Beyond cross-validation or conventional validation within a dataset, external replication in a separate cohort is ideal – and even further value can be gained from using different metabolomics platforms.[35] In addition, confirmatory results from performing different types of statistical analyses (e.g., traditional and non-traditional) in both training and validation cohorts may be informative.

**Conclusions**

High-throughput metabolomics data provide an exciting area of research for scientific discovery, but are accompanied by a number of statistical challenges that must be properly addressed to infer meaningful results from these complex data. Herein, we have reviewed and outlined practical solutions for many of the common problems found in metabolomics data analysis. While outlining some guidelines for future researchers on how to address these issues, the optimal solution at any given stage of data management and analysis depends on the size and design of the specific data and study at hand. Thus, a critical aspect of any analysis is explicit recognition of the strengths and weaknesses of the selected approaches as well as a complete understanding of the assumptions that go into each decision that is made throughout the analysis workflow. Many important decisions, such as handling missing data and transforming right-skewed data, can have meaningful impacts on the final analysis results. Therefore, we recommend that researchers perform sensitivity analyses with respect to these decisions to assess the robustness of their results to such subjective choices. In addition, any findings from primary analyses should be replicated in additional studies to confirm their potential to serve as meaningful scientific findings, irrespective of the statistical decisions made.

While we have reviewed potential solutions for the problems that are often encountered in human metabolomics studies, many of these issues do not have definitive answers, and this presents possibilities for future methods research that can aim to improve decision making at each stage of the analysis workflow. Certain pervasive challenges, such as how to best handle missing data, may be addressed using a number of possible approaches although most of these approaches are ad hoc and often favored due to their ease of application. The same can be said for the transformation of variables as well as how to address batch-to-batch variability across a large cohort-wide experiment. These are all open areas of research, and finding optimal solutions will lead to substantial improvements in the analysis, reproducibility, and overall contributions in the field of human metabolomics investigation.

**COMPETING INTERESTS**

The authors declare no conflicts of interest.

**CONTRIBUTIONS**

Designed the study: JA, BC, MH, JDW, KL, MJ, and SC. Conducted the experiments: JA, BC, MH, JDW, and KL. Analysed the data: MH, TN, JDW, KL, BVJ, MH, and SC. Provided material, data or analysis tools: JA, BC, MH. Wrote the paper: JA, BC, MJ, SC. All authors read the paper and contributed to its final form.

**FUNDING**

This project was supported in part by the National Institutes of Health grants T32-ES007142 (JA), R01-HL134168 (SC, MJ), K01-HL135342 (OD), R01-HL134811 (OD, SM), the American Heart Association (CVGPS Pathway Award: MGL, SC, MJ), the Doris Duke Charitable Foundation (MJ, SC), and the Tobacco Related Disease Research Program (MJ, JDW).

**Table 1. Statistical Considerations for Human Metabolomics Data**

| Consideration | Notes and Examples |
|---|---|
| Missingness | • Patterns of missing values tend to be non-random and are even sometimes predictable. For example, missing values may often but not always be more frequent for metabolites that are intrinsically low in abundance when measured from a given tissue type.<br>• Missingness may be due to biological and/or technical reasons. |
| Data distributions | • Many but not all metabolites tend to demonstrate right-skewed distributions in most types of human studies (e.g., healthy controls or disease-specific referral samples).<br>• Certain metabolites will display a substantial proportion of zero values that may be considered true zero values based on biology (an issue to be considered along with but distinguished from missingness). |
| Intercorrelations | • Intercorrelations between metabolites may well reflect clustering of small molecules by known or (mostly) unknown biological pathways.<br>• Intercorrelations will vary widely depending on a given exposure or background, chronic disease status, and other yet unidentified factors.<br>• Intercorrelations will also vary depending on the underlying MS method used to create a given dataset (i.e., untargeted vs. targeted, and the specific technical methods used). |
| Time-dependence | • Whereas a portion of the human metabolome changes dynamically in response to acute perturbation or stress, many other metabolites display variation only over several days to weeks in response to subacute perturbations; other portions of the metabolome may yet exhibit relatively little change over time, except in response to major chronic exposures. |
| Confounding factors | • Metabolite values will vary in response to factors that are measurable as well as factors that are not easily measurable for a given study, such as acute and chronic dietary patterns, microbiota, and environmental exposures. |

Table 2. Statistical Analysis Methods for Outcomes Analyses of Human Metabolomics Data

| Method | Univariate or Multivariate | Handling binary outcome | Handling continuous outcome | P value for significant metabolites | Metabolite selection | Advantages | Disadvantages |
|---|---|---|---|---|---|---|---|
| Multiple tests (e.g., univariate linear regression) with Bonferroni correction | Univariate | Yes | Yes | Yes | Yes | Simple, easy to use and interpret results | Very conservative and does not account for intercorrelation |
| Multiple tests with false discovery rate (FDR) | Univariate | Yes | Yes | Yes | Yes | Simple, easy to use, less conservative than Bonferroni correction | |
| Principal component analysis (PCA) | Multivariate | Yes | Yes | No | No | Effective for variable reduction | No intrinsic clarity on how to select or rank variables |
| Sparse partial least squares (SPLS) | Multivariate | No | Yes | No | Yes | | |
| Linear discriminant analysis (LDA) | | Yes | No | No | Yes | | |
| Random forests and other machine learning approaches | Multivariate | Yes | Yes | No | No | Can find complex relationships between variables | If data is truly linear, this will be less efficient |
| Least absolute shrinkage and selection operator (LASSO) | Multivariate | Yes | Yes | No | Yes | | |

**Figure 1. Metabolite data transformation and centering.** A frequently used approach for managing metabolite data collected in a large human cohort study involves log transforming each metabolite measures and centering the data on plate median to account for batch to batch variation.

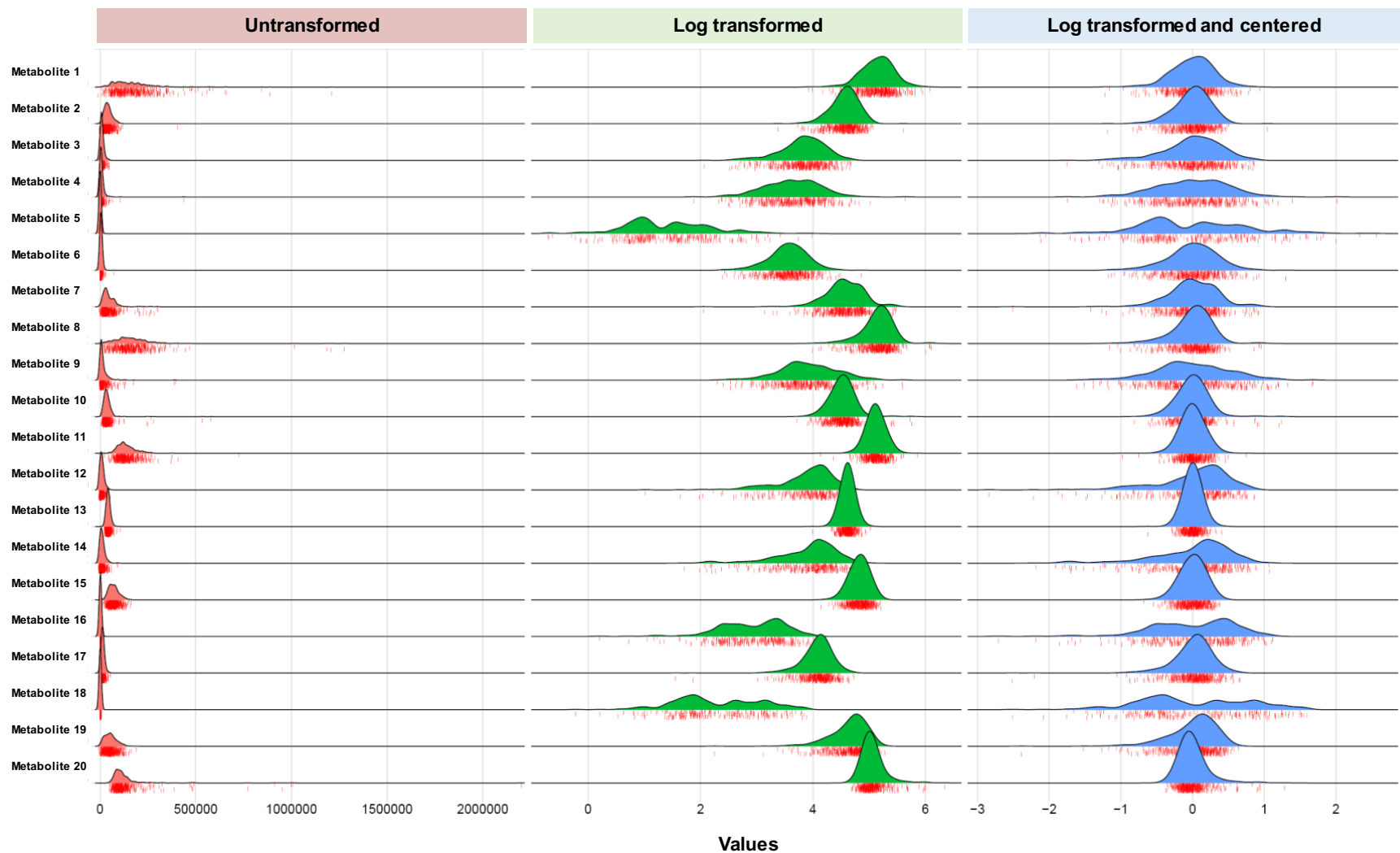

**Figure 2. Actual and simulated metabolomics data.** Previously analyzed data, or prior detailed knowledge of the structure of metabolomics data collected from an existing human cohort study (Panel A) can be used to construct simulated data that mimics the data structure observed from real measures (Panel B). These simulated data can be used to estimate statistical power, based on one or more methods of analyses, for planning the design of a future study.

**A.**

**B.**

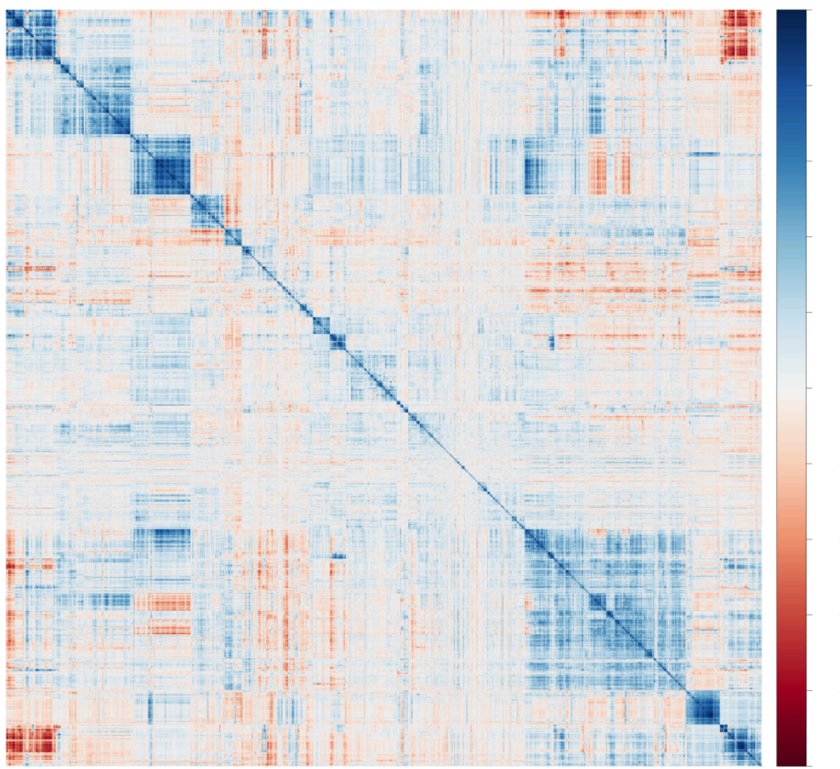
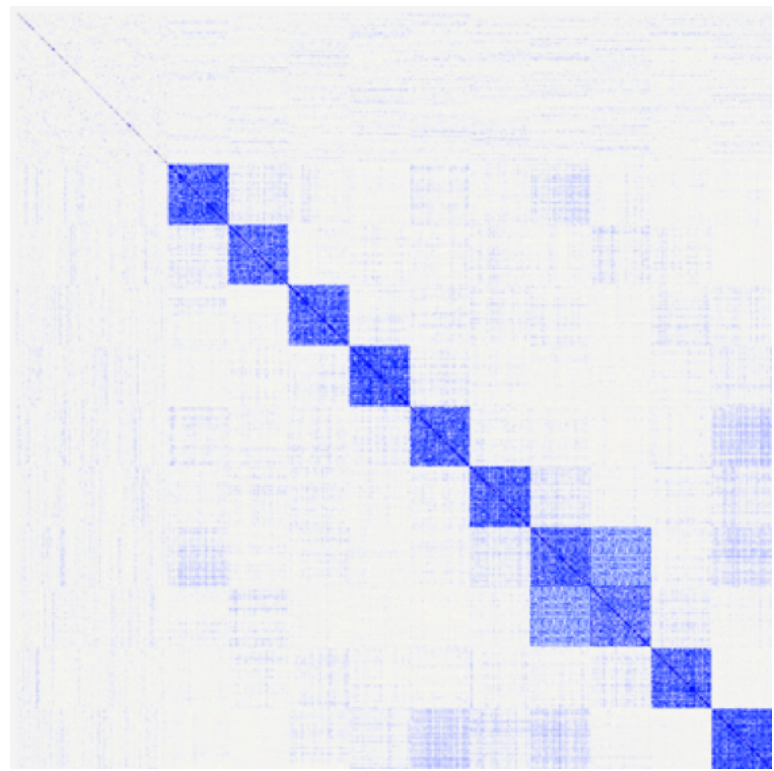

**Figure 3. Using multiple statistical methods to evaluate results.** When relating a panel of metabolites to a given outcome, multiple different statistical methods can be used with the results compared. Although conventional and machine learning models tend to agree on the top ranked metabolite associations with a given outcome, more divergent and potentially complementary information is offered by the second and third tier metabolite associations. Discordant results likely reflect different assumptions and features between methods, such as the assumption of linearity of association between a predictor and outcome for conventional regression models. FDR, false discovery rate; LASSO, least absolute shrinkage and selection operator; RF, random forests; MCFS, monte carlo feature selection; RKNN, random K nearest neighbors.

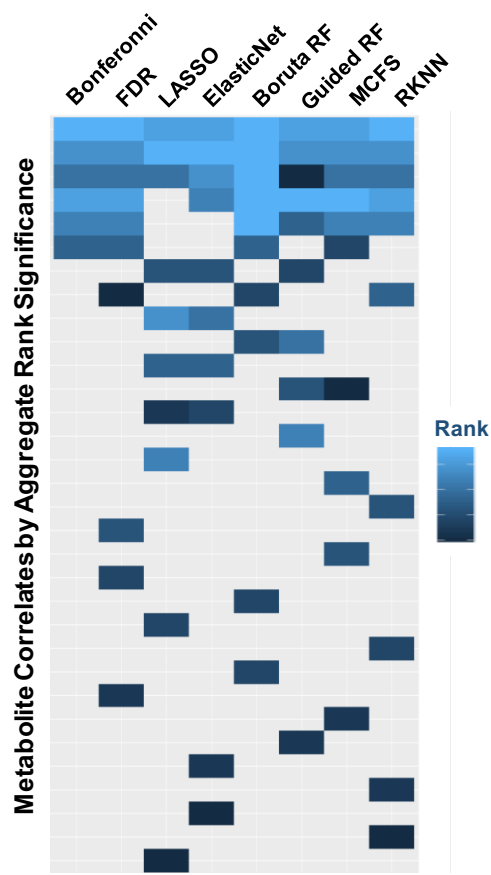

**Figure 4. Visualizing results.** Different approaches to visualizing the results of relating multiple metabolites with multiple clinical outcomes in a large human metabolomics study are possible, including but not limited to combined Manhattan plots (Panel A) and a paired heatmap depicting values for both beta coefficients and P values (Panel B).

A.

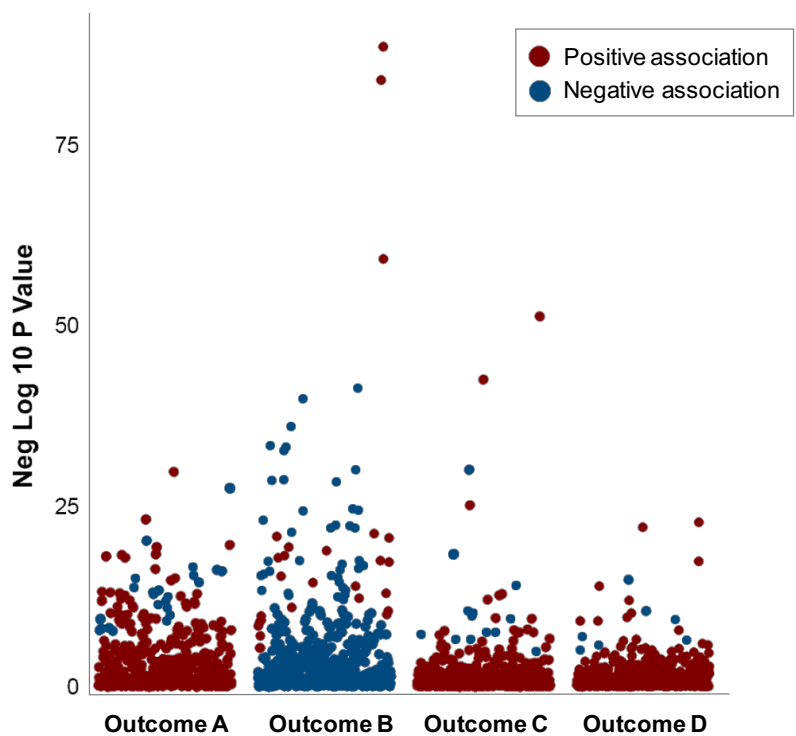

B.

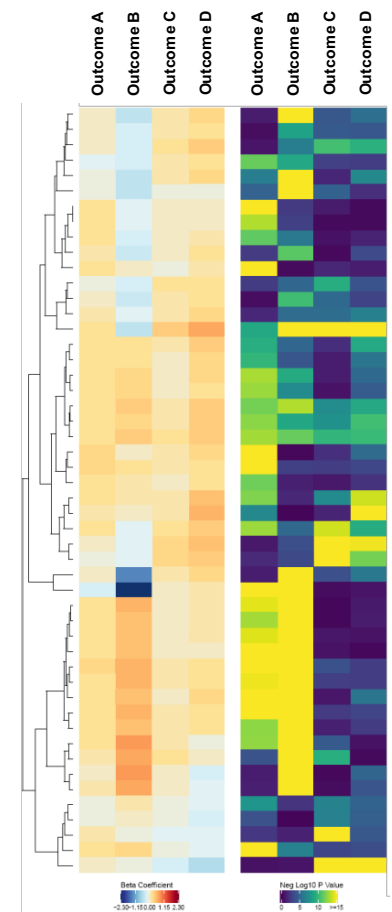